\documentclass[10pt,notitlepage,a4paper]{article}
\usepackage{axodraw}
\usepackage[dvips]{graphicx}

\newcommand{\ie}{{\it i.e.}}
\newcommand{\eg}{{\it e.g.}}

\newcommand{\Pbf}{{\bf P}}

\newcommand{\pbf}{{\bf p}}
\newcommand{\kbf}{{\bf k}}
\newcommand{\xbf}{{\bf x}}
\newcommand{\qbf}{{\bf q}}
\newcommand{\gambf}{{\mathbf{\gamma}}}

\newcommand{\ieps}{i\epsilon}

\newcommand{\lsim}{\buildrel < \over {_\sim}}
\newcommand{\aslash}[1]{ \rlap{/}{#1} }

\newcommand{\order}[1]{${\cal O}\left(#1 \right)$}
\newcommand{\morder}[1]{{\cal O}\left(#1 \right)}

\newcommand{\beq}{\begin{equation}}
\newcommand{\eeq}{\end{equation}}
\newcommand{\beqa}{\begin{eqnarray}}
\newcommand{\eeqa}{\end{eqnarray}}
\newcommand{\beqat}{\begin{eqnarray*}}
\newcommand{\eeqat}{\end{eqnarray*}}
\newcommand{\ket}[1]{\vert{#1}\rangle}
\newcommand{\bra}[1]{\langle{#1}\vert}

\newcommand{\PL}[3]{Phys.~Lett.~{\bf {#1}},~{#2}~({#3})}
\newcommand{\NP}[3]{Nucl.~Phys.~{\bf {#1}},~{#2}~({#3})}

\newcommand{\PRD}[3]{Phys.~Rev.~{\bf D{#1}},~{#2}~({#3})}

\newcommand{\PR}[3]{Phys.~Rev.~{\bf {#1}},~{#2}~({#3})}
\newcommand{\tr}{\mathrm{Tr}}
\newcommand{\trs}[1]{\tr{\left\{ {#1} \right\}}}
\newcommand{\prm}{\textrm{ .}}
\newcommand{\crm}{\textrm{ ,}}

\newcommand{\gz}{\gamma^0}

\newcommand{\ud}{\textrm{d}}

\newcommand{\Tim}[1]{\mathrm{T}\left\{ {#1} \right\}}

\renewcommand{\thefootnote}{\fnsymbol{footnote}}

\begin{document}

 \begin{flushright}
   hep-ph/0411208\\
   HIP-2004-61/TH
 \end{flushright}

 \begin{centering}
  {\LARGE Hydrogen Atom in Relativistic Motion}

  \vspace{12pt}

  {\large M. J\"arvinen\footnote{Email address: Matti.O.Jarvinen@Helsinki.fi}$^,$\footnote{Research supported in part by the 
   Academy of Finland through grant 102046.}}

  \vspace{7pt}

  {\large Department of Physical Sciences and \\ \vspace{1pt} Helsinki Institute of Physics\\
   \vspace{2pt} POB 64, FIN-00014 University of Helsinki, Finland}

  \vspace{3pt}

 \end{centering}

\setcounter{footnote}{0}
\renewcommand{\thefootnote}{\arabic{footnote}}

\section*{Abstract}

The Lorentz contraction of bound states in field theory is often
appealed to in qualitative descriptions of high energy particle
collisions. Surprisingly, the contraction has not been demonstrated
explicitly even in simple cases such as the hydrogen atom. 
It requires a calculation of wave functions evaluated at equal (ordinary) time for
bound states in motion. Such wave functions are not obtained by
kinematic boosts from the rest frame. Starting from the exact
Bethe-Salpeter equation we derive the equal-time wave function of
a fermion-antifermion bound state in QED, \ie, positronium or the 
hydrogen atom, in any frame to leading order in $\alpha$. We show
explicitly that the bound state energy transforms as the fourth
component of a vector and that the wave function of the 
fermion-antifermion Fock state contracts as expected.
Transverse photon exchange contributes at leading order to the binding
energy of the bound state in motion. We study the general features of the
corresponding fermion-antifermion-photon Fock states, and show that they
do not transform by simply contracting.
We verify that the wave function reduces to the light-front one in the infinite momentum frame.

\section{Introduction}

Bound state wave functions are usually considered only in their
center-of-mass frame, where rotational symmetry may be fully exploited.
In the study of scattering amplitudes involving several bound states
one needs, however, to know the wave functions in arbitrary Lorentz
frames. Center-of-mass wave functions are commonly defined at equal
time $t=0$ of the constituents, and are non-trivially related to equal
time wave functions in motion since the relative time is  boost
dependent \cite{Brodsky}. Consequently, the Hamiltonian does not
commute with the boost generators, causing the boost to be as
complicated as solving the bound state equation directly in the new
frame.

One manifestation of the non-trivial boost dynamics is the expectation,
based on classical relativistic physics, that the equal-time wave
function of a bound state in motion will be contracted in the direction
of motion. High energy hadron scattering is thus often pictured as
collisions between Lorentz-contracted pancakes. This is necessarily a
qualitative description since we are far from being able to calculate
the wave function of a hadron even in the center-of-mass frame.
However, it is surprising that the equal-time wave functions of much
simpler bound states, such as the hydrogen atom, have apparently only
been considered in the rest frame.

The hydrogen atom is a non-relativistic state (at leading order in the
fine structure constant $\alpha$) and the calculation of its wave
function is one of the first exercises in courses on quantum mechanics.
When the atom is in relativistic motion we must, however, make use of
the full machinery of relativistic field theory. The excitation
energies of Fock states with additional particles (electron-positron
pairs, or photons) may be much less than the energy of the bound state.
Hence the contributions of such higher Fock states must be carefully
considered. As we shall see below, the exchange of transverse photons
indeed contributes at leading order to the binding of hydrogen in
motion. This is not unexpected, as a boost of the Coulomb potential
leads to a transverse electromagnetic field.

The Lorentz contraction of 3+1 dimensional wave functions in gauge
field theories has apparently not been demonstrated previously. In
\cite{Brodsky} a Lorentz contracting wave function of a two body QED
bound state is represented as an approximation valid for small boosts.
The frame dependence of bound state wave functions has been studied in
various
models, see for example, \cite{Gloeckle,Hoyer,Schon}. In \cite{Fukui,Guinea} Lorentz
contraction is obtained for a fermion pair interacting via a
$\delta$ potential. There has also been other interesting work on the
Lorentz covariance of two body equations \cite{Hanson,Artru}.

It is reasonable to expect that the equal-time wave function of the
hydrogen atom (or positronium) can be evaluated analytically in any
frame to leading order in $\alpha$. We find that the wave function of
the $e^+ e^-$ Fock state indeed contracts as expected from classical
relativity. The probability of the $e^+ e^- \gamma$ state is of
\order{\alpha}, which reflects the relative scarceness of photon
exchange in the weak coupling limit. The photon amplitude does not
classically contract, however. More generally, quantum fluctuations are
unlikely to obey classical transformation laws.

Rather than trying to boost the well-known rest frame wave function of
positronium we time order and solve its Bethe-Salpeter equation
\cite{Bethe} for an arbitrary momentum of the bound state. We
generalize our previous 1+1 dimensional calculation \cite{Jarv} by
including the transverse photons which contribute in 3+1 dimensions.
The equal-time formalism is of necessity Lorentz non-covariant --
nevertheless we shall see that the bound state energy transforms as the
fourth component of a Lorentz vector. We study the properties of the
transverse photon distribution and show that it agrees with the
light-front result in the limit of an infinitely large bound state
momentum.

\section{Wave equation for the hydrogen atom in motion}

We calculate here the equal $x^0$ wave function for a (3+1 dimensional) fermion-antifermion bound state in the weak coupling limit $\alpha \ll 1$ of QED in
any Lorentz frame. We will follow closely the procedure of \cite{Jarv} in 1+1 dimensions: the Bethe-Salpeter bound state equation
is solved to leading order in
$\alpha$ but to all orders in $|\Pbf|/m$. Here $\Pbf$ is the total momentum of the system and the fermion masses $m$ are taken to be equal
for notational simplicity. We use Coulomb gauge (${\bf k}\cdot{\bf A}=0)$, but also check that the result holds in Feynman gauge. In Coulomb gauge the unphysical photon polarization states are absent and the contribution from physical, transverse polarizations is best seen.

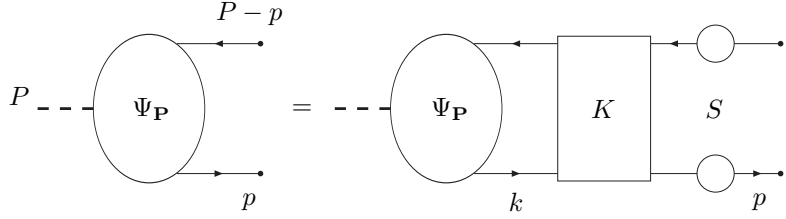
\begin{figure}
 \SetScale{0.7}\begin{picture}(300,80)(-33,0)
  \Oval(70,50)(40,30)(0)
  \SetWidth{1.5}
  \DashLine(10,50)(40,50){7}
  \SetWidth{0.5}
  \ArrowLine(130,85)(85,85)
  \ArrowLine(85,15)(130,15)
  \Vertex(130,15){1.5}
  \Vertex(130,85){1.5}
  \Text(107,35)[]{$=$}
  \Oval(230,50)(40,30)(0)
  \SetWidth{1.5}
  \DashLine(170,50)(200,50){7}
  \SetWidth{0.5}
  \ArrowLine(290,85)(245,85)
  \ArrowLine(245,15)(290,15)
  \Boxc(315,50)(50,78)
  \ArrowLine(365,85)(340,85)
  \Oval(375,85)(10,10)(0)
  \Line(385,85)(410,85)
  \ArrowLine(385,15)(410,15)
  \Oval(375,15)(10,10)(0)
  \Line(340,15)(365,15)
  \Vertex(410,15){1.5}
  \Vertex(410,85){1.5}
  \Text(221,35)[]{$K$}
  \Text(263,35)[]{$S$}
  \Text(50,35)[]{$\Psi_{\Pbf}$}
  \Text(163,35)[]{$\Psi_{\Pbf}$}
  \Text(0,40)[]{$P$}
  \Text(87,0)[]{$p$}
  \Text(87,71)[]{$P-p$}
  \Text(280,0)[]{$p$}
  \Text(188,0)[]{$k$}
 \end{picture}
 \caption{The Bethe-Salpeter equation. The blobs represent the wave function $\Psi_{\Pbf}$, $K$ is the interaction kernel and $S$ is the two-particle propagator.}
 \label{waveeq}
\end{figure}

Our starting point, the Bethe-Salpeter equation, is defined as shown in Fig. \ref{waveeq}.
The propagator $S$ is the summed propagator including all radiative corrections.
The interaction kernel $K$ includes all two-particle irreducible interaction diagrams, \ie,
diagrams that cannot be split into two interaction graphs just by cutting two fermion lines. It is amputated such that
it does not include the outgoing or incoming fermion propagators which are included in $S$.
The Bethe-Salpeter wave function $\Psi_\Pbf$
is defined as the projection of the bound state onto a fermion-antifermion state
\beq \label{wfdef}
 \Psi_{\Pbf}(p)_{\alpha\beta} = \int \ud^4 x\,e^{ix\cdot p} \bra{\Omega}\Tim{\bar\psi_\beta(0)\psi_\alpha(x)}\ket{\Pbf \lambda}
\eeq
where $\Pbf$ is the total momentum, $p$ is the four momentum of the fermion, $\bra{\Omega}$ is the vacuum of QED and $\lambda$ represents all discrete quantum numbers of the bound state $\ket{\Pbf \lambda}$
(such as spin and orbital angular momentum).

As in \cite{Jarv}, we work in a time-ordered formalism where the Fock space structure of $K$ and $\Psi_\Pbf$ is seen explicitly.
The time-ordered rules are obtained by taking a Fourier transform over $p^0$. The fermion propagator has forward and backward moving parts in ($t,\pbf$)
space
\beqa \label{fermprop}
 S_F(t,\pbf) &\equiv& \int \frac{\ud p^0}{2\pi}\exp(-it p^0)\, \frac{i}{\aslash p-m+\ieps} \nonumber\\
  &=& \Theta(t)\ \Lambda^+(\pbf) \exp(-it E_\pbf) + \Theta(-t)\ \Lambda^-(-\pbf) \exp(it E_\pbf)
\eeqa
where $E_\pbf = \sqrt{\pbf^2+m^2}$ and the projection operators $\Lambda^\pm$ are defined by
\beq \label{Lambdadef}
 \Lambda^\pm(\pbf) \equiv \frac{\pm\gz E_\pbf\mp\gambf\cdot\pbf+m}{2E_\pbf} \prm
\eeq
Similarly, the photon propagator in ($t,\pbf$) space reads in Feynman gauge
\beq \label{fgpp}
 D_F^{\mu\nu}(t,\pbf) = -\frac{g^{\mu\nu}}{2|\pbf|}\left[\Theta(t)\exp(-i|\pbf|t) + \Theta(-t)\exp(i|\pbf|t)\right] \prm
\eeq
In Coulomb gauge ($\pbf\cdot{\bf A}=0$) we have an instantaneous contribution from the $D^{00}$ component
\beqa
  D_C^{00}(t,\pbf) &=& \delta(t)\frac{i}{\pbf^2}\textrm{ ;}\quad\quad D_C^{i0}=D_C^{0i} = 0 \nonumber\\
  D_C^{ij}(t,\pbf) &=& \left(\delta^{ij}-\frac{p^i p^j}{\pbf^2}\right)\frac{1}{2|\pbf|}\left[\Theta(t)\exp(-i|\pbf|t) + \Theta(-t)\exp(i|\pbf|t)\right] \prm
\eeqa
In the usual time-ordered perturbation theory one integrates over all time differences from zero to infinity. The integrals give
energy denominators which are denoted by vertical cuts in the Feynman diagrams in the following sections.

In the center-of-mass frame of positronium the scale of the internal momenta and the scale of the binding energy or the potential energy are
\beq \label{rfscales}
 |\pbf| \sim \alpha m\crm\quad \Delta E \sim V \sim \alpha^2 m \crm
\eeq
respectively. For a moving system we define the relative momentum by
\beq \label{intmom}
 \qbf \equiv \pbf - \Pbf/2 \prm
\eeq
We expect the transverse components of $\qbf$ and of the photon momentum $\kbf$ to be the same order as in the rest frame (\ref{rfscales}),
whereas the longitudinal components and the energy differences will be affected by the contraction,
\beq \label{scales}
 q_\parallel \sim k_\parallel \sim \gamma \alpha m\crm\quad |\qbf_\perp| \sim |\kbf_\perp| \sim \alpha m\crm \quad \Delta E \sim \gamma^{-1} \alpha^2 m 
\eeq
where $\gamma\equiv \sqrt{\Pbf^2+(2m)^2}/2m$ is the boost parameter of the bound state (to lowest order in $\alpha$).

\subsection{Structure of the interaction kernel}

\label{strucsec}

Next we will identify the leading contributions to $K$ using time-ordered diagrams in the weak coupling limit. We will see that they arise from the
single photon exchange part of $K$ which involves only Fock states with one additional photon. 
The radiative corrections to the fermion propagators which are included in $S$ of Fig. \ref{waveeq} renormalize the mass and change the off-shell
dependence of the propagator. In a non-relativistic bound state the constituents are nearly on-shell. Hence to leading order in $\alpha$ these
effects are accounted for by using the physical mass $m$ in the propagator (\ref{fermprop}). 

When iterated the Bethe-Salpeter equation (Fig. \ref{waveeq})
gives the wave function as an infinite ladder diagram with rungs composed of the kernel $K$. 
We will time order the ladder and analyze its blocks.
We work here in Coulomb gauge, but it is easy to check that the results are
gauge independent. 

\begin{figure}
\SetScale{0.7}\begin{picture}(140,100)(-30,-10)
 \ArrowLine(125,80)(70,80)
 \ArrowLine(70,20)(125,20)
 \DashLine(125,20)(125,80){7}
 \ArrowLine(125,20)(180,20)
 \ArrowLine(180,80)(125,80)
 \DashLine(145,5)(145,95){3}
 \Text(55,25)[]{$\pbf-\kbf$}
 \Text(55,65)[]{$\Pbf-\pbf+\kbf$}
 \Text(122,25)[]{$\pbf$}
 \Text(122,65)[]{$\Pbf-\pbf$}
 \LongArrow(120,55)(120,45)
 \Text(80,37)[]{$\kbf$}

 \ArrowLine(280,80)(250,80)
 \ArrowLine(250,20)(280,20)
 \DashLine(300,5)(300,95){3}
 \Photon(280,80)(320,20){3}{5}
 \Line(280,20)(320,20)
 \Line(280,80)(320,80)
 \DashLine(335,5)(335,95){3}
 \ArrowLine(360,80)(320,80)
 \ArrowLine(320,20)(360,20)
 \Text(181,25)[]{$\pbf-\kbf$}
 \Text(181,65)[]{$\Pbf-\pbf+\kbf$}
 \Text(248,25)[]{$\pbf$}
 \Text(255,65)[]{$\Pbf-\pbf$}
 \Text(208,78)[]{$\Delta E_I$}
 \Text(236.5,78)[]{$\Delta E_F$}
 \LongArrow(307,53)(315,42)
 \Text(225,38)[]{$\kbf$}

 \Text(90,-10)[]{(a)}
 \Text(216,-10)[]{(b)}

\end{picture}

\caption{Time-ordered single photon exchange diagrams which arise from the interaction kernel $K$ of Fig. \ref{waveeq}. (a) The instantaneous Coulomb interaction 
$K_\gamma^a$. (b) The exchange of a transverse photon $K_\gamma^b$. Time flows to the right.}
\label{opfsd}
\end{figure}
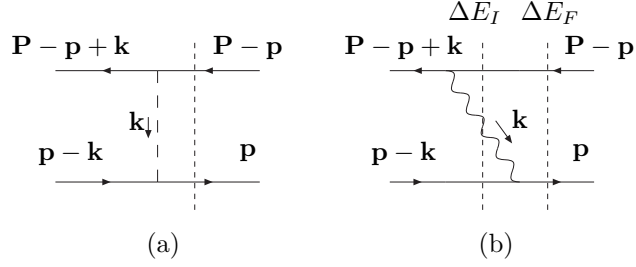

Let us first analyze the single photon exchange diagrams (Fig. \ref{opfsd}). Using time-ordered Feynman rules in Coulomb gauge, we
have
\beqa \label{opfsc}
 K_\gamma^a &=& \frac{i}{E-E_\pbf-E_{\Pbf-\pbf}}\int \frac{\ud^3 \kbf}{(2 \pi)^3} \frac{i}{\kbf^2}\nonumber\\
 &&\cdot \Lambda^+(\pbf)ie\gz\Lambda^+(\pbf-\kbf) \otimes \Lambda^-(\Pbf-\pbf+\kbf) ie\gz\Lambda^-(\Pbf-\pbf) \nonumber\\
 K_\gamma^b &=& \frac{i}{E-E_\pbf-E_{\Pbf-\pbf}}\int \frac{\ud^3 \kbf}{(2 \pi)^3} \frac{1}{2|\kbf|}\frac{i}{E-E_{\pbf-\kbf}-E_{\Pbf-\pbf}-|\kbf|} \\ \nonumber
 &&\cdot\ \left(\delta^{ij}-\frac{k^ik^j}{\kbf^2}\right)\Lambda^+(\pbf)ie\gamma^i\Lambda^+(\pbf-\kbf) \otimes \Lambda^-(\Pbf-\pbf+\kbf)ie\gamma^j\Lambda^-(\Pbf-\pbf)
\eeqa
where $\otimes$ denotes a direct product between the Dirac spaces of the fermions. For the energy denominators arising from the cuts of Fig. \ref{opfsd}
we have [using (\ref{intmom}) and (\ref{scales})]\footnote{We will see that the scaling behavior $\gamma^{-1}$ of
$\Delta E_F$ is an exact result whereas $\Delta E_I \propto \gamma^{-1}$ holds for $k_\parallel>0$. For $k_\parallel<0$ we would have $\Delta E_I \sim \gamma
\alpha m$. This reflects the fact that for $\gamma \gg 1$ backward moving photons are suppressed, see Sec. \ref{photonsec}.}
\beqa \label{endiff}
\Delta E_F &\equiv& E-E_\pbf-E_{\Pbf-\pbf} \sim \Delta E\ \sim\ \alpha^2 m \gamma^{-1}\\
\Delta E_I &\equiv& E-E_{\pbf-\kbf}-E_{\Pbf-\pbf}-|\kbf| = E-E_\pbf-E_{\Pbf-\pbf} + \left(E_{\pbf}-E_{\pbf-\kbf}-|\kbf|\right) \nonumber\\
 &\sim& \left(E_{\pbf}-E_{\pbf-\kbf}-|\kbf|\right) \sim\ \alpha m \gamma^{-1} \prm \label{gendiff}
\eeqa

From (\ref{endiff}) we have for the energy differences $\Delta E_F\simeq \alpha \Delta E_I$: for the excitation of one photon we need an energy \order{1/\alpha}
higher than the binding energy in all frames. 
Correspondingly, for the time scales $\Delta t_{I/F}\equiv 1/\Delta E_{I/F}$ we have $\Delta t_I \simeq \alpha \Delta t_F$.  
Hence transverse photon exchange is a rare event
in the weak coupling limit: The probability of finding the bound state in an excited Fock state should be the ratio
of the two time scales and thus $\sim \alpha$. This expectation will be confirmed in Sec. \ref{confirm}.
\label{expect}

In Fig. \ref{opfsd} and in the above analysis we assumed that all the fermions move forward in time. However, the fermion propagator (\ref{fermprop}) also contains a backward moving component. Its inclusion leads to ``Z graphs'' involving pair production which are suppressed in the non-relativistic
limit due to large energy denominators.

Let us pay attention to the coupling structure appearing in (\ref{opfsc}).
Recalling the definition of $\Lambda^\pm$ in (\ref{Lambdadef}), we use the property
$\{\gamma^\mu,\gamma^\nu\}=2g^{\mu\nu}$ to write
\beq \label{Lambdaexp}
 \Lambda^+(\pbf-\kbf)\gamma^\mu\Lambda^+(\pbf) = \Lambda^+(\pbf-\kbf)\left[\frac{p^\mu}{E_\pbf} + \Lambda^-(\pbf)\gamma^\mu \right] \prm
\eeq
Using $\Lambda^+(\pbf)\Lambda^-(\pbf)=0$ and the results (\ref{intmom}), (\ref{scales}) we can estimate
\beqa \label{estims}
 \Lambda^+(\pbf-\kbf)\Lambda^-(\pbf) &=& \morder{|\kbf|/\mathcal{E}} = \morder{\alpha} \nonumber\\
 \frac{p^\mu}{E_\pbf} &=& \frac{P^\mu}{\mathcal{E}} + \morder{\alpha}
\eeqa
where $P^0\simeq\mathcal{E}\equiv\sqrt{(2m)^2+\Pbf^2}$. Inserting (\ref{estims}) into (\ref{Lambdaexp}) we have
\beq \label{cest}
 \Lambda^+(\pbf-\kbf)\gamma^\mu\Lambda^+(\pbf) = \Lambda^+(\pbf-\kbf)\frac{P^\mu}{\mathcal{E}} + \morder{\alpha} \prm
\eeq
A similar analysis for the antifermion coupling reveals that at leading order we may replace
\beqa \label{crepl}
 \gamma^\mu \Lambda^+(\pbf) &\longrightarrow& \frac{P^\mu}{\mathcal{E}} \nonumber\\
 \Lambda^-(\Pbf-\pbf)\gamma^\mu &\longrightarrow& -\frac{P^\mu}{\mathcal{E}} \prm
\eeqa

Using the above estimates we can see that in a general frame both diagrams of Fig. \ref{opfsd} are of the
same order in $\alpha$. Let us first analyze diagram (a). The first factor in Eq. (\ref{opfsc}) gives $\Delta t_F \sim \gamma \alpha^{-2}m^{-1}$. Using (\ref{crepl}) we have
\beq \label{cstruct}
  \Lambda^+(\pbf)\gamma^0 \otimes \gamma^0 \Lambda^-(\Pbf-\pbf) \sim \left(\frac{P^0}{\mathcal{E}}\right)^2 \simeq 1 \prm
\eeq
The Coulomb potential gives using (\ref{scales})
\beq
  \alpha \int \frac{\ud^3 \kbf}{(2\pi)^3} \frac{1}{\kbf^2} \sim \alpha\frac{\kbf_\perp^2 k_\parallel}{|\kbf|^2} \sim \gamma^{-1}\alpha^2m \equiv V\prm 
\eeq
Altogether
\beq \label{opfsa}
 K_\gamma^a \sim \Delta t_F \cdot V \sim \gamma \alpha^{-2}m^{-1} \cdot \gamma^{-1}\alpha^2 m = \alpha^0
\eeq
where we dropped the $k$ dependent projectors $\Lambda^+(\pbf-\kbf)$ and $\Lambda^-(\Pbf-\pbf+\kbf)$ which belong to a different block of $K\cdot S$ (see Fig. \ref{waveeq}).
For diagram (b)
\beq \label{trphc}
 \left( \delta^{ij}-\frac{k^ik^j}{\kbf^2} \right)  \Lambda^+(\pbf)\gamma^i \otimes \gamma^j \Lambda^-(\Pbf-\pbf) \sim 
\left( \delta^{ij}-\frac{k^ik^j}{\kbf^2} \right) \frac{P^i}{\mathcal{E}}\frac{P^j}{\mathcal{E}}= \beta^2 \frac{\kbf_\perp^2}{\kbf^2}\sim\frac{\beta^2}{\gamma^2}
\eeq
where $\beta\equiv|\Pbf|/\mathcal{E}$. The potential contributes
\beq
  \alpha \int \frac{\ud^3 \kbf}{(2\pi)^3} \frac{1}{2|\kbf|}\frac{i}{E-E_\pbf-E_{\Pbf-\pbf+\kbf}-|\kbf|}  \sim 
\alpha \Delta t_I \frac{\kbf_\perp^2 k_\parallel}{|\kbf|} \sim \gamma\alpha^2m 
\eeq
and thus
\beq \label{opfsab}
 K_\gamma^b \sim \gamma \alpha^{-2}m^{-1} \cdot \beta^2\gamma^{-2} \cdot\gamma\alpha^2m =\beta^2\alpha^0 \prm
\eeq

In particular, due to the couplings (\ref{trphc}), the contribution from transverse photons [Fig. \ref{opfsd} (b)]
is absent at order $\alpha^0$ in the center-of-mass frame ($\beta=0$).
This is the expected result: for the hydrogen atom at rest transverse photons do not contribute at leading order, but
appear as spin dependent interactions at next-to-leading order. The spin dependent interactions are hidden in the $\morder{\alpha}$
terms of (\ref{cest}) also when $\Pbf \ne 0$.

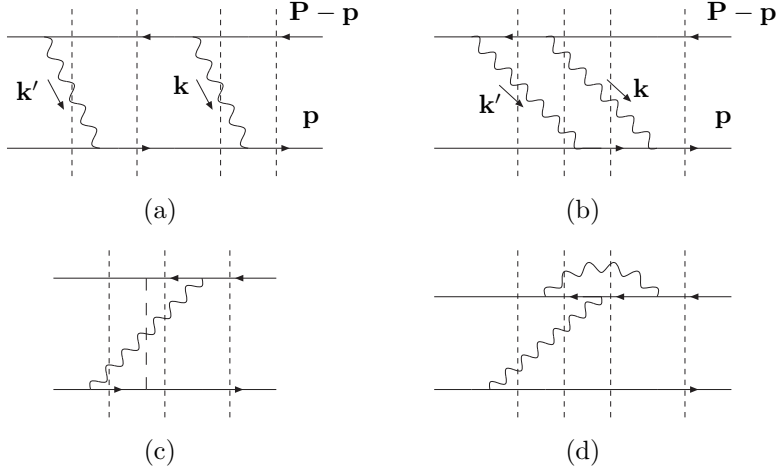
\begin{figure}
\SetScale{0.7}\begin{picture}(140,180)(-30,-105)
 \Line(10,80)(70,80)
 \Line(10,20)(70,20)
 \Photon(60,20)(30,80){3}{5}
 \ArrowLine(70,20)(100,20)
 \ArrowLine(100,80)(70,80)
 \Line(100,80)(140,80)
 \Line(100,20)(140,20)
 \Photon(140,20)(110,80){3}{5}
 \ArrowLine(140,20)(180,20)
 \ArrowLine(180,80)(140,80)
 \DashLine(45,5)(45,95){3}
 \DashLine(80,5)(80,95){3}
 \DashLine(125,5)(125,95){3}
 \DashLine(155,5)(155,95){3}
 \Text(122,25)[]{$\pbf$}
 \Text(127,65)[]{$\Pbf-\pbf$}
 \LongArrow(112,57)(120,42)
 \LongArrow(32,57)(40,42)
 \Text(15,36)[]{$\kbf'$}
 \Text(73,38)[]{$\kbf$}

 \Line(330,20)(240,20)
 \Line(240,80)(260,80)
 \DashLine(285,5)(285,95){3}
 \Photon(320,20)(260,80){3}{7}
 \Photon(360,20)(300,80){3}{7}
 \ArrowLine(300,80)(260,80)
 \DashLine(310,5)(310,95){3}
 \DashLine(335,5)(335,95){3}
 \DashLine(375,5)(375,95){3}
 \ArrowLine(320,20)(360,20)
 \Line(300,80)(360,80)
 \ArrowLine(400,80)(360,80)
 \ArrowLine(360,20)(400,20)
 \Text(278,25)[]{$\pbf$}
 \Text(285,65)[]{$\Pbf-\pbf$}
 \LongArrow(275,54)(287,43)
 \LongArrow(333,57)(345,46)
 \Text(190,31)[]{$\kbf'$}
 \Text(247,37)[]{$\kbf$}

 \Line(35,-50)(85,-50)
 \Line(35,-110)(55,-110)
 \ArrowLine(55,-110)(85,-110)
 \ArrowLine(115,-50)(85,-50)
 \Line(85,-110)(115,-110)
 \DashLine(85,-110)(85,-50){7}
 \Photon(55,-110)(115,-50){3}{7}
 \ArrowLine(115,-110)(155,-110)
 \ArrowLine(155,-50)(115,-50)
 \DashLine(65,-125)(65,-35){3}
 \DashLine(95,-125)(95,-35){3}
 \DashLine(130,-125)(130,-35){3}

 \Line(300,-60)(240,-60)
 \ArrowLine(330,-60)(300,-60)
 \Line(240,-110)(260,-110)
 \DashLine(285,-125)(285,-35){3}
 \Photon(270,-110)(330,-60){3}{7}
 \PhotonArc(330,-80)(36.06,33.7,146.3){3}{5.5}
 \Line(260,-110)(300,-110)
 \DashLine(310,-125)(310,-35){3}
 \DashLine(335,-125)(335,-35){3}
 \DashLine(375,-125)(375,-35){3}
 \ArrowLine(360,-60)(320,-60)
 \Line(300,-110)(360,-110)
 \ArrowLine(400,-60)(360,-60)
 \ArrowLine(360,-110)(400,-110)

 \Text(65,-10)[]{(a)}
 \Text(225,-10)[]{(b)}
 \Text(65,-101)[]{(c)}
 \Text(225,-101)[]{(d)}

\end{picture}

\caption{Time-ordered two photon exchange diagrams.
Diagrams (a) and (b) arise from iterating the covariant Bethe-Salpeter equation (Fig. \ref{waveeq}) with one photon exchange kernels. Diagrams (c) and
(d) arise from two photon kernels.
}
\label{tpfsd}
\end{figure}

Next we will show that more complicated diagrams can be neglected in all frames.
A representative set of two photon exchange diagrams is shown in Fig. \ref{tpfsd}.
Diagram (a) will be included in our approximation. Diagrams (b) and (d) which include Fock states
with two photons and diagram (c) will be suppressed.

Let us study more closely diagrams (a) and (b). Diagram (a) simply consists of two separate transverse photon exchanges
of Fig. \ref{opfsd}(a), and we have from (\ref{opfsa})
\beq
 K_{\gamma\gamma}^a \sim (K_\gamma^b)^2 \sim \beta^4\alpha^0 \prm
\eeq
However, in diagram (b) three of the cuts intersect photon lines instead of two.
Quantitatively, the only difference to diagram (a) comes from the second cut from the left:
\beqa
 (a):&\quad&\frac{1}{E-E_{\pbf-\kbf}-E_{\Pbf-\pbf+\kbf}} \sim \gamma\alpha^{-2}m^{-1} =\Delta t_F \nonumber\\
 (b):&\quad&\frac{1}{E-E_{\pbf}-E_{\Pbf-\pbf+\kbf+\kbf'}-|\kbf|-|\kbf'|} \lsim \gamma\alpha^{-1}m^{-1} = \Delta t_I \prm
\eeqa
That is, in diagram (a) the two interactions are separated by the long time scale $\Delta t_F$, but in diagram
(b) both the interactions must occur within the shorter time scale $\Delta t_I$. We have
\beq
 K_{\gamma\gamma}^b \sim \frac{\Delta t_I}{\Delta t_F} K_{\gamma\gamma}^a \sim \beta^4\alpha
\eeq
and the diagram is thus suppressed. 
The qualitative picture is that the flight time of a photon is short compared to the intervals between the excanges.
Thus the probability of having two photons at the same instant of time is low. 
Similar arguments show that diagrams (c) and (d) are also suppressed.

In the case of positronium the kernel $K$ includes annihilation diagrams with, \eg,
one photon as an intermediate Fock state. Similar arguments as above show that these graphs are \order{\alpha^2}
and thus suppressed in all frames.

We assumed in this section that momenta of order $|\kbf_\perp| \sim \alpha m$ (and $k_\parallel \sim \gamma \alpha m$) dominate the integrations, which is true at leading order in $\alpha$. Transverse photons with softer momenta 
$|\kbf_\perp| \sim \alpha^2 m$ are suppressed by \order{\alpha} due to the smaller phase space. However,
the flight time of such photons is comparable with the longer time scale $\Delta t_F$. This allows an arbitrary number of harder 
$|\kbf_\perp| \sim \alpha m$ (transverse or Coulomb) exchanges
while the soft photon is in flight. As the harder interactions are \order{\alpha^0} contributions, the diagrams similar to the one shown in Fig. \ref{tpfsd}(c) but with, \eg, several Coulomb exchanges are in fact all \order{\alpha}. Such diagrams are known to contribute to bound state structure at higher orders in the center-of-mass frame \cite{Salpeter}.

\subsection{Lorentz contraction of the e$^+$e$^-$ wave function}

\begin{figure}
\centering
\SetScale{0.7}\begin{picture}(300,155)(20,-85)
 \Oval(90,50)(40,30)(0)
 \SetWidth{1.5}
 \DashLine(30,50)(60,50){7}
 \SetWidth{0.5}
 \ArrowLine(150,85)(105,85)
 \ArrowLine(105,15)(150,15)
 \Vertex(150,15){1.5}
 \Vertex(150,85){1.5}
 \Text(63,35)[]{$\varphi_\Pbf$}

 \Text(160,35)[]{=}

 \Oval(310,50)(40,30)(0)
 \SetWidth{1.5}
 \DashLine(250,50)(280,50){7}
 \SetWidth{0.5}
 \ArrowLine(370,85)(325,85)
 \ArrowLine(325,15)(370,15)
 \DashLine(370,15)(370,85){7}
 \ArrowLine(370,15)(415,15)
 \ArrowLine(415,85)(370,85)
 \Vertex(415,85){1.5}
 \Vertex(415,15){1.5}
 \DashLine(388,5)(388,95){3}
 \Text(33,24)[]{$\Pbf$}
 \Text(96,18)[]{$\pbf$}
 \LongArrow(380,60)(380,40)
 \Text(275,35)[]{$\kbf$}
 \Text(290,0)[]{$\pbf$}
 \Text(250,0)[]{$\pbf-\kbf$}
 \Text(217,35)[]{$\varphi_\Pbf$}

 \Text(301,35)[]{+}

 \Oval(90,-70)(40,30)(0)
 \SetWidth{1.5}
 \DashLine(30,-70)(60,-70){7}
 \SetWidth{0.5}
 \ArrowLine(130,-35)(105,-35)
 \ArrowLine(105,-105)(170,-105)
 \Photon(130,-35)(170,-105){-3}{5}
 \ArrowLine(200,-35)(130,-35)
 \ArrowLine(170,-105)(200,-105)
 \Vertex(200,-35){1.5}
 \Vertex(200,-105){1.5}
 \DashLine(150,-25)(150,-115){3}
 \DashLine(180,-25)(180,-115){3}
 \Text(115,-46)[]{$\kbf$}
 \Text(140,-85)[]{$\pbf$}
 \Text(90,-85)[]{$\pbf-\kbf$}
 \LongArrow(152,-60)(162,-80)
 \Text(63,-49)[]{$\varphi_\Pbf$}

 \Text(160,-49)[]{+}

 \Oval(310,-70)(40,30)(0)
 \SetWidth{1.5}
 \DashLine(250,-70)(280,-70){7}
 \SetWidth{0.5}
 \ArrowLine(390,-35)(325,-35)
 \ArrowLine(325,-105)(350,-105)
 \Photon(390,-35)(350,-105){3}{5}
 \ArrowLine(420,-35)(390,-35)
 \ArrowLine(350,-105)(420,-105)
 \Vertex(420,-35){1.5}
 \Vertex(420,-105){1.5}
 \DashLine(370,-25)(370,-115){3}
 \DashLine(400,-25)(400,-115){3}
 \Text(248,-46)[]{$\kbf$}
 \Text(294,-85)[]{$\pbf$}
 \Text(234,-85)[]{$\pbf-\kbf$}
 \LongArrow(368,-60)(358,-80)
 \Text(217,-49)[]{$\varphi_\Pbf$}

 \end{picture}
\caption{The time-ordered bound state equation in the ladder approximation. The blobs denote the equal-time wave function $\varphi_{\Pbf}$ defined in (\ref{etwfdef}).}
\label{tobse}
\end{figure}
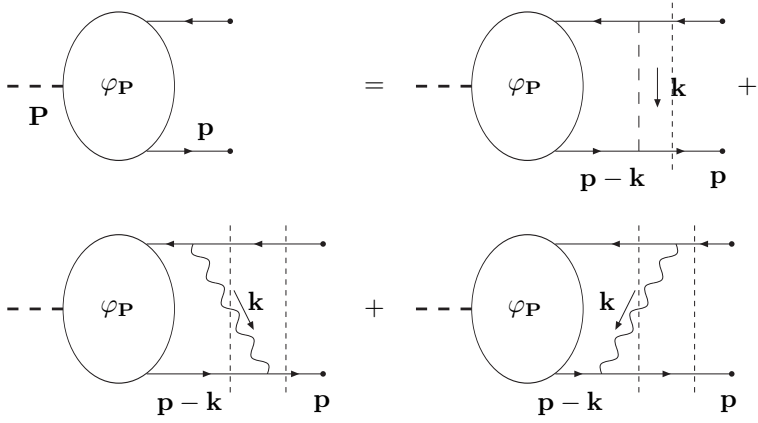

Now we are ready to write the bound state equation (Fig. \ref{waveeq}) in time-ordered form with only the leading diagrams included. 
After the Fourier transform over energy, the relevant wave function is the equal-time wave function
\beq \label{etwfdef}
 \varphi_\Pbf(\pbf)_{\alpha\beta} \equiv \int \frac{\ud p^0}{2\pi} \Psi_\Pbf(p)_{\alpha\beta} \prm
\eeq
The equation is shown in Fig. \ref{tobse}. The analytic expression reads
\beqa \label{bsest}
 \varphi_\Pbf(\pbf) &=& \frac{i}{E-E_\pbf-E_{\Pbf-\pbf}} \int \frac{\ud^3 \kbf}{(2\pi)^3}\Bigg[\frac{i}{\kbf^2} \Lambda^+(\pbf)\,ie\gz \varphi_\Pbf(\pbf-\kbf)\, ie\gz \Lambda^-(\Pbf-\pbf) \nonumber\\
 &+& \left(\frac{i}{E-E_{\pbf-\kbf}-E_{\Pbf-\pbf}-|\kbf|} + \frac{i}{E-E_\pbf-E_{\Pbf-\pbf+\kbf}-|\kbf|}\right) \nonumber\\
 &\cdot& \frac{1}{2|\kbf|}\left(\delta^{ij} -\frac{k^ik^j}{\kbf^2}\right) \Lambda^+(\pbf)\,ie\gamma^i \varphi_\Pbf(\pbf-\kbf)\, ie\gamma^j \Lambda^-(\Pbf-\pbf)\Bigg]
\eeqa
where $\varphi_\Pbf$ is understood as a 4x4 Dirac matrix. In particular, from the property $\Lambda^\pm\Lambda^\mp =0 $ of the projection
matrices (\ref{Lambdadef}) it follows that
\beq \label{formov}
 \Lambda^-(\pbf)\varphi_\Pbf(\pbf)=0=\varphi_\Pbf(\pbf)\Lambda^+(\Pbf-\pbf) \crm
\eeq
\ie, the wave function (\ref{etwfdef}) has only forward moving components in the weak coupling limit.

At leading order in $\alpha$ we may use the replacement (\ref{crepl}) to eliminate the Dirac structure. We have
\beqa \label{bsean}
 \varphi_\Pbf(\pbf) &=& \frac{-e^2}{E-E_\pbf-E_{\Pbf-\pbf}} \int \frac{\ud^3 \kbf}{(2 \pi)^3}\Bigg[\frac{1}{\kbf^2} +
 \frac{1}{\mathcal E^2}\left(\Pbf^2-\frac{(\Pbf\cdot\kbf)^2}{\kbf^2}\right)\frac{1}{2|\kbf|} \\ \nonumber
 &\cdot&\left(\frac{1}{E-E_\pbf-E_{\Pbf-\pbf+\kbf}-|\kbf|} + \frac{1}{E-E_{\pbf-\kbf}-E_{\Pbf-\pbf}-|\kbf|} \right) \Bigg]\varphi_\Pbf(\pbf-\kbf)\\
 &\equiv& \frac{1}{E-E_\pbf-E_{\Pbf-\pbf}}\int \frac{\ud^3 \kbf}{(2 \pi)^3} V(\kbf) \varphi_\Pbf(\pbf-\kbf) \label{vdef}
\eeqa
where $\mathcal{E}\equiv\sqrt{\Pbf^2+(2m)^2}$.
The equation thus reduces to a scalar equation for the forward moving components
of $\varphi_\Pbf$. In comparison with the 1+1 dimensional equation of Ref. \cite{Jarv}, we now have a contribution from transverse photon exchange.

Let us study the sum $E_\pbf+E_{\Pbf-\pbf}$ appearing in the denominator of (\ref{vdef}). Expanding in the relative momentum $\qbf$ of (\ref{intmom}) we get
\beq \label{ensum}
 E_\pbf+E_{\Pbf-\pbf} - \mathcal E = \frac{1}{2\mu\gamma}\left(\qbf_\perp^2 + \gamma^{-2} q_\parallel^2\right) + \morder{\gamma^{-1}\alpha^4 m}
\eeq
where $\gamma\equiv \mathcal{E}/2m$ and $\mu\equiv m/2$. Note that the form of (\ref{ensum}) is consistent with our expectation (\ref{scales}).
For the energy denominators in the transverse part of (\ref{bsean}) we get
\beqa \label{endiffest}
 E-E_{\pbf-\kbf}-E_{\Pbf-\pbf}-|\kbf| &=& \left(E-E_\pbf-E_{\Pbf-\pbf}\right)+\left(E_\pbf-E_{\pbf-\kbf}-|\kbf|\right) \nonumber\\
   &=& \frac{\Pbf\cdot\kbf}{\mathcal E}-|\kbf|+\morder{\gamma^{-1}\alpha^2m} \nonumber\\
E-E_\pbf-E_{\Pbf-\pbf+\kbf}-|\kbf| &=& -\frac{\Pbf\cdot\kbf}{\mathcal E}-|\kbf|+\morder{\gamma^{-1}\alpha^2m}
\eeqa
and the potential defined in (\ref{vdef}) becomes
\beqa \label{kpot}
 \frac{1}{4\pi\alpha}V(\kbf)&=&- \frac{1}{\kbf^2} + \frac{\beta^2\kbf_\perp^2}{2 \kbf^2}\left(\frac{1}{\kbf^2 + \beta k_\parallel|\kbf|} + \frac{1}{\kbf^2 - \beta k_\parallel|\kbf|} \right) \\
&=&- \frac{1}{\kbf^2} + \frac{\beta^2\kbf_\perp^2}{\kbf^2\left(\kbf_\perp^2+\gamma^{-2} k_\parallel^2\right)} 
 = - \frac{1}{\gamma^2(\kbf_\perp^2+\gamma^{-2} k_\parallel^2)} \label{kpotfin}
\eeqa
in terms of the parallel and perpendicular components of $\kbf$ and with $\beta\equiv|\Pbf|/\mathcal E$. In particular, the contribution from the transverse
photons [second term in (\ref{kpot})] is proportional to $\beta^2$ and vanishes in the center-of-mass frame. Requiring the potential energy (\ref{kpotfin}) 
to be commensurate with the relative kinetic energy in (\ref{ensum}) finally verifies our expectation (\ref{scales}) and the $\gamma$ dependence of the non-leading terms in (\ref{ensum}), (\ref{endiffest}).

Let us define the binding energy $\Delta M \equiv \sqrt{E^2-\Pbf^2}-2m$ which should be independent of $\Pbf$. As $E-\mathcal E = \Delta E \sim \gamma^{-1}\alpha^2m$, we have
\beq \label{massa}
 \Delta M = \gamma(E-\mathcal E) + \morder{\alpha^4 m} \prm
\eeq
Inserting (\ref{ensum}), (\ref{kpotfin}) and (\ref{massa}) into (\ref{vdef}) we get
\beq \label{fffswe}
 \left[ \Delta M - \frac{1}{2\mu}\left(\qbf_\perp^2 + \gamma^{-2} q_\parallel^2\right)\right] \varphi_\Pbf(\pbf)
 = - \frac{4 \pi \alpha}{\gamma} \int \frac{\ud^3\kbf}{(2\pi)^3} \frac{\varphi_\Pbf(\pbf-\kbf)}{\kbf_\perp^2+\gamma^{-2} k_\parallel^2} \prm
\eeq
We see that all frame dependence, \ie, the $\gamma$ factors, can be removed by rescaling $k_\parallel \rightarrow \gamma k_\parallel$ and 
$q_\parallel \rightarrow \gamma q_\parallel$.
Then the spectrum is given by the same Schr\"odinger equation as in the center-of-mass frame.
The e$^+$e$^-$ wave function exactly Lorentz contracts (or expands in $\kbf$ space) in the direction of motion.
Taking into account the Dirac structure (\ref{formov}),
the result can be written
\beq \label{Dirs}
 \varphi_\Pbf(\pbf)_{\alpha\beta} = \sum_{s_1,s_2}\frac{u_\alpha(\Pbf/2,s_1)\bar v_\beta(\Pbf/2,s_2)}{2E_{\Pbf/2}} \chi_{s_1,s_2}
 \phi_\Pbf( \qbf)
\eeq
where $u$, $v$ are the usual Dirac spinors and $\phi_\Pbf$ is given by Lorentz contracting the usual wave function $\phi_{CM}$ of the hydrogen atom at rest:
\beq \label{hydrwf}
 \phi_\Pbf(\qbf) = \frac{1}{\sqrt{\gamma}} \phi_{CM}(\qbf_\perp,q_\parallel/\gamma) \prm
\eeq
The spin wave function $\chi$ is normalized as $\sum_{s_1,s_2}|\chi_{s_1,s_2}|^2=1$.
At leading order in $\alpha$ we were able to replace $\pbf$ and $\Pbf-\pbf$ by $\Pbf/2$ in the Dirac structure of
(\ref{Dirs}).

We finally check that the result (\ref{Dirs}) is also obtained in Feynman gauge with the propagator (\ref{fgpp}). In this gauge the photon propagator
has no instantaneous part and (\ref{bsest}) becomes
\beqa
 \varphi_\Pbf(\pbf) &=& \frac{i}{E-E_\pbf-E_{\Pbf-\pbf}} \int \frac{\ud^3 \kbf}{(2\pi)^3}
 \Lambda^+(\pbf)\,ie\gamma^\mu \varphi_\Pbf(\pbf-\kbf)\, ie\gamma_\mu \Lambda^-(\Pbf-\pbf) \nonumber\\
 &\cdot& \frac{-1}{2|\kbf|} \left(\frac{i}{E-E_{\pbf-\kbf}-E_{\Pbf-\pbf}-|\kbf|} + \frac{i}{E-E_\pbf-E_{\Pbf-\pbf+\kbf}-|\kbf|}\right) \prm
\eeqa
Using (\ref{crepl}), the Dirac structure reduces to a factor $-\gamma^{-2}$. Further using (\ref{endiffest}), the potential in Feynman gauge becomes
\beq
 \frac{1}{4\pi\alpha}V(\kbf)=-\frac{1}{\gamma^2}\cdot\frac 12\left(\frac{1}{\kbf^2 + \beta k_\parallel|\kbf|} + \frac{1}{\kbf^2 - \beta k_\parallel|\kbf|} \right)
 = -\frac{1}{\gamma^2(\kbf_\perp^2+\gamma^{-2} k_\parallel^2)}
\eeq
which is the same result as in Coulomb gauge (\ref{kpotfin}). The rest of the calculation remains unchanged.

\section{The e$^+$e$^-\gamma$ wave function}
\label{photonsec}

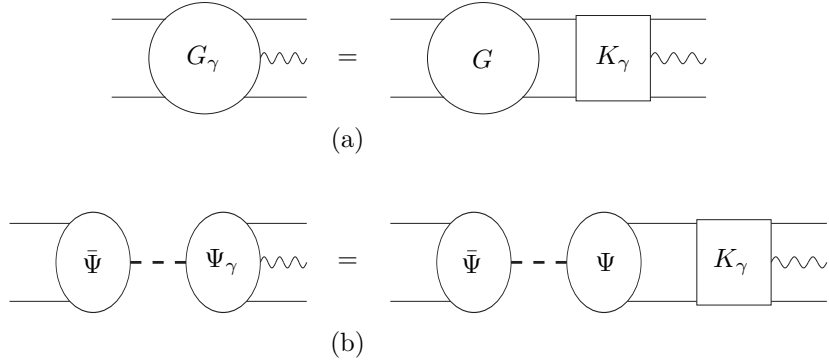
\begin{figure}
 \SetScale{0.7}\begin{picture}(300,160)(-20,-90)
   \Line(94,71)(65,71)
   \Line(94,29)(65,29)
   \Oval(115,50)(30,30)(0)
   \Line(136,71)(170,71)
   \Line(136,29)(170,29)
   \Photon(145,50)(170,50){3}{3}
   \Text(80.5,35)[]{$G_\gamma$}
   \Text(135,35)[]{$=$}
   \Line(244,71)(215,71)
   \Line(244,29)(215,29)
   \Oval(265,50)(30,30)(0)
   \Line(286,71)(315,71)
   \Line(286,29)(315,29)
   \Text(186,35)[]{$G$}
   \Boxc(335,50)(40,46)
   \Text(235.5,35)[]{$K_\gamma$}
   \Line(385,71)(355,71)
   \Line(385,29)(355,29)
   \Photon(355,50)(385,50){3}{3}

  \Text(135,4)[]{(a)}

   \Line(10,-39)(42.4,-39)
   \Line(10,-81)(42.4,-81)
   \Oval(55,-60)(27,20)(0)
   \Text(38.5,-42)[]{$\bar \Psi$}
   \SetWidth{1.5}
   \DashLine(75,-60)(105,-60){7}
   \SetWidth{0.5}
   \Oval(125,-60)(27,20)(0)
   \Line(137.6,-39)(170,-39)
   \Line(137.6,-81)(170,-81)
   \Photon(145,-60)(170,-60){3}{3}
   \Text(87.5,-42)[]{$\Psi_\gamma$}
   \Text(135,-42)[]{$=$}
   \Line(215,-39)(247.4,-39)
   \Line(247.4,-81)(215,-81)
   \Oval(260,-60)(27,20)(0)
   \Text(182,-42)[]{$\bar \Psi$}
   \SetWidth{1.5}
   \DashLine(280,-60)(310,-60){7}
   \SetWidth{0.5}
   \Oval(330,-60)(27,20)(0)
   \Line(342.6,-39)(380,-39)
   \Line(342.6,-81)(380,-81)
   \Text(232.5,-42)[]{$\Psi$}
   \Boxc(400,-60)(40,46)
   \Text(280,-42)[]{$K_\gamma$}
   \Line(450,-39)(420,-39)
   \Line(450,-81)(420,-81)
   \Photon(420,-60)(450,-60){3}{3}

   \Text(135,-73)[]{(b)}

 \end{picture}
 \caption{Derivation of the Bethe-Salpeter equation for the photon wave function. (a) The generating equation for the five-point Green function. $G$ is the
 four-fermion Green function, $G_\gamma$ has in addition a photon insertion and $K_\gamma$ is the two-particle irreducible kernel.
 (b) The pole contribution giving the Bethe-Salpeter equation.
  Arrows on the quark lines were dropped for simplicity.}
 \label{bsder}
\end{figure}

In the preceding section we derived the wave function for the e$^+$e$^-$ Fock state, which is
expected to be the leading component of the bound state in the weak coupling limit. Now we will analyze the wave function of the
e$^+$e$^-\gamma$ Fock state: we calculate the distribution of physical, transverse photons in the bound state. 
We will see that this state occurs with a probability
\order{\alpha}, whereas Fock states with more photons or e$^+$e$^-$ pairs contribute terms of \order{\alpha^2} to the normalization.  

The e$^+$e$^-\gamma$ Fock state contributes as an intermediate state in the derivation of (\ref{fffswe}), see Fig. \ref{tobse}.
However, to clarify the derivation of its wavefunction it is best to start again from the exact covariant
formalism: we will briefly repeat the analysis of the previous section but now for the e$^+$e$^-\gamma$ Fock state.
The photon distribution will be related to the square of the equal-time wave function of the Fock state just as in usual non-relativistic
quantum mechanics. For an analogous calculation on the light-front see \cite{Burkardt,Gunion}.

Let us define the wave function as a coupling to an e$^+$e$^-\gamma$ state
\beq
 \Psi_{\gamma\,\Pbf}(p,k) \equiv \int \ud^4 x\ud^4 y\,e^{+ix\cdot (p-k) +i y \cdot k} \bra{\Omega}\Tim{\bar\psi_\beta(0)\psi_\alpha(x) A^\mu(y)}
 \ket{\Pbf \lambda} \prm
\eeq
A covariant equation analogous to the Bethe-Salpeter equation connecting $\Psi_\gamma$ to the wave function $\Psi$ of (\ref{wfdef})
may be derived in a way analogous to the derivation of the original Bethe-Salpeter equation (see \cite{Bethe,Nakanishi}).
We only sketch the proof here (see Fig. \ref{bsder}). The five-point Green function $G_\gamma$ satisfies the identity shown in Fig. \ref{bsder}a which
may be proved diagrammatically. The kernel $K_\gamma$ is defined to be the sum of all two-particle irreducible diagrams, \ie, diagrams
which cannot be divided into two separate diagrams by cutting two fermion lines. The wave equation is then obtained by calculating
the residues at the pole caused by the bound state (see Fig. \ref{bsder}b), \ie, at $P^2=M^2$, where $P$ is the total momentum and
$M$ is the bound state mass. $\Psi_\gamma$ is simply given by adding the kernel $K_\gamma$ to $\Psi$.
Note that $K_\gamma$ cannot have a pole at $P^2=M^2$: this would lead to $G_\gamma$ having a double pole.

\begin{figure}
\SetScale{0.7}\begin{picture}(140,175)(-10,-100)

 \ArrowLine(135,80)(45,80)
 \Line(45,20)(90,20)
 \ArrowLine(90,20)(135,20)
 \DashLine(108,5)(108,95){3}
 \PhotonArc(135,-60)(110,90,133.3){3}{8}

 \Line(230,80)(180,80)
 \Photon(200,20)(260,80){3}{7}
 \Line(180,20)(200,20)
 \DashLine(210,5)(210,95){3}
 \DashLine(235,5)(235,95){3}
 \DashLine(275,5)(275,95){3}
 \ArrowLine(260,80)(220,80)
 \Line(200,20)(260,20)
 \ArrowLine(300,80)(260,80)
 \ArrowLine(260,20)(300,20)
 \PhotonArc(300,160)(110,226.7,270){-3}{8}

 \Line(380,80)(330,80)
 \Photon(350,20)(410,80){3}{7}
 \Line(330,20)(350,20)
 \DashLine(360,5)(360,95){3}
 \DashLine(385,5)(385,95){3}
 \DashLine(425,5)(425,95){3}
 \Line(410,80)(370,80)
 \ArrowLine(350,20)(370,20)
 \Line(370,20)(410,20)
 \ArrowLine(450,80)(410,80)
 \ArrowLine(410,20)(450,20)
 \PhotonArc(450,-60)(110,90,133.3){3}{8}

 \Line(45,-50)(95,-50)
 \Line(45,-110)(65,-110)
 \ArrowLine(65,-110)(95,-110)
 \DashLine(95,-110)(95,-50){7}
 \ArrowLine(95,-110)(135,-110)
 \ArrowLine(135,-50)(95,-50)
 \DashLine(75,-125)(75,-35){3}
 \DashLine(110,-125)(110,-35){3}
 \PhotonArc(135,-190)(110,90,133.3){-3}{8}

 \Line(200,-60)(180,-60)
 \ArrowLine(230,-60)(200,-60)
 \PhotonArc(230,-80)(36.06,33.7,146.3){3}{5.5}
 \Line(180,-110)(200,-110)
 \DashLine(210,-125)(210,-35){3}
 \DashLine(235,-125)(235,-35){3}
 \DashLine(275,-125)(275,-35){3}
 \ArrowLine(260,-60)(220,-60)
 \Line(200,-110)(260,-110)
 \ArrowLine(300,-60)(260,-60)
 \ArrowLine(260,-110)(300,-110)
 \PhotonArc(300,20)(110,226.7,270){-3}{8}

 \Line(320,-50)(330,-50)
 \Line(320,-110)(345,-110)
 \Photon(330,-50)(360,-70){-3}{4}
 \Photon(345,-110)(375,-90){3}{4}
 \Line(360,-70)(375,-90)
 \ArrowLine(430,-50)(330,-50)
 \ArrowLine(345,-110)(460,-110)
 \ArrowLine(400,-70)(360,-70)
 \ArrowLine(375,-90)(415,-90)
 \ArrowLine(430,-50)(460,-50)
 \Line(400,-70)(415,-90)
 \Photon(430,-50)(400,-70){3}{4}
 \Photon(415,-90)(460,-90){-3}{4}
 \DashLine(337.5,-125)(337.5,-35){3}
 \DashLine(352.5,-125)(352.5,-35){3}
 \DashLine(367.5,-125)(367.5,-35){3}
 \DashLine(387.5,-125)(387.5,-35){3}
 \DashLine(407.5,-125)(407.5,-35){3}
 \DashLine(422.5,-125)(422.5,-35){3}
 \DashLine(442.5,-125)(442.5,-35){3}

 \Text(63,-8)[]{(a)}
 \Text(168,-8)[]{(b)}
 \Text(273,-10)[]{(c)}
 \Text(63,-99)[]{(d)}
 \Text(168,-99)[]{(e)}
 \Text(273,-99)[]{(f)}

\end{picture}

\caption{Time-ordered diagrams arising from the interaction kernel $K_\gamma$.
(a) is a leading diagram in $\alpha$.
}
\label{tospk}
\end{figure}
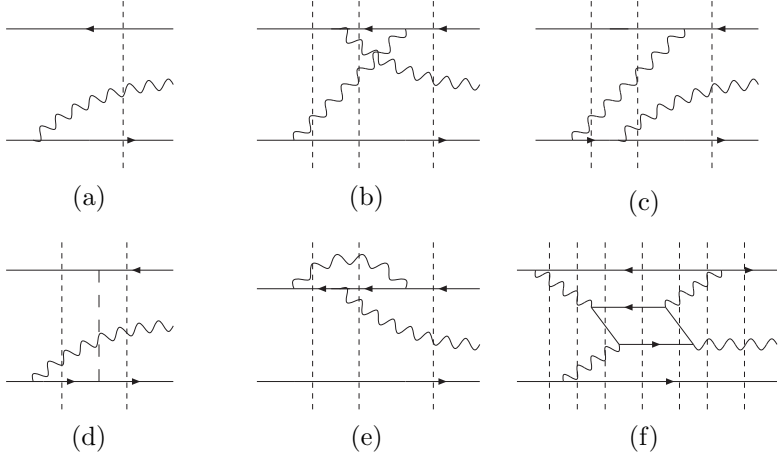

Let us move to the time-ordered formalism. The equal-time wave function for the single photon Fock state may be defined as
\beqa \label{photonwf}
 \Phi^\mu_\Pbf(\pbf,\kbf) &\equiv&\int \frac{\ud p^0}{2\pi}\frac{\ud k^0}{2\pi} \Psi_{\gamma\,\Pbf}(p,k) \nonumber\\
 &=&\int \ud^3 {\bf x}\ud^3{\bf y}\,e^{-i{\bf x}\cdot(\pbf-\kbf) -i{\bf y}\cdot\kbf} \bra{\Omega}\bar\psi_\beta(0)\psi_\alpha(x) A^\mu(y)\ket{\Pbf \lambda}\Big|_{x^0=y^0=0} \nonumber\\
&\equiv& \bra{\Omega} \bar \psi_\beta(0)\widetilde \psi_\alpha(\pbf-\kbf)\widetilde A^\mu(\kbf) \ket{\Pbf \lambda} \prm
\eeqa
When time ordered, the interaction $K_\gamma$ gives rise to several graphs of which some are shown in Fig. \ref{tospk}.
Diagram (c) arises in fact as a combination of the lowest order kernels $K$ and $K_\gamma$, but will be suppressed 
since the two photons overlap in time.

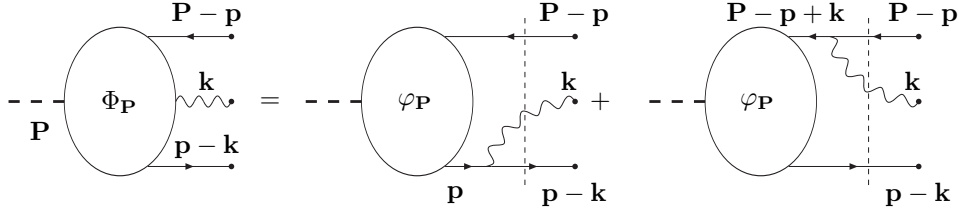
\begin{figure}
\centering
\SetScale{0.7}\begin{picture}(300,75)(46,0)
 \Oval(90,50)(40,30)(0)
 \SetWidth{1.5}
 \DashLine(30,50)(60,50){7}
 \SetWidth{0.5}
 \ArrowLine(150,85)(105,85)
 \ArrowLine(105,15)(150,15)
 \Vertex(150,15){1.5}
 \Vertex(150,85){1.5}
 \Vertex(150,50){1.5}
 \Photon(120,50)(150,50){3}{3}
 \Text(33,24)[]{$\Pbf$}
 \Text(96,18)[]{$\pbf-\kbf$}
 \Text(96,43)[]{$\kbf$}
 \Text(96,68)[]{$\Pbf-\pbf$}
 \Text(63,35)[]{$\Phi_\Pbf$}

 \Text(120,35)[]{=}

 \Oval(250,50)(40,30)(0)
 \SetWidth{1.5}
 \DashLine(190,50)(220,50){7}
 \SetWidth{0.5}
 \ArrowLine(335,85)(265,85)
 \ArrowLine(265,15)(290,15)
 \ArrowLine(290,15)(335,15)
 \Vertex(335,85){1.5}
 \Vertex(335,15){1.5}
 \Vertex(335,50){1.5}
 \DashLine(308,5)(308,95){3}
 \PhotonArc(335,0)(50,90,162.5){3}{5}
 \Text(175,35)[]{$\varphi_\Pbf$}
 \Text(235,0)[]{$\pbf-\kbf$}
 \Text(233,43)[]{$\kbf$}
 \Text(235,68)[]{$\Pbf-\pbf$}
 \Text(190,0)[]{$\pbf$}

 \Text(245,35)[]{+}

 \Oval(435,50)(40,30)(0)
 \SetWidth{1.5}
 \DashLine(375,50)(405,50){7}
 \SetWidth{0.5}
 \ArrowLine(475,85)(450,85)
 \ArrowLine(520,85)(475,85)
 \ArrowLine(450,15)(520,15)
 \Vertex(520,85){1.5}
 \Vertex(520,15){1.5}
 \Vertex(520,50){1.5}
 \DashLine(493,5)(493,95){3}
 \PhotonArc(520,100)(50,197.5,270){-3}{5}
 \Text(304.5,35)[]{$\varphi_\Pbf$}
 \Text(365,0)[]{$\pbf-\kbf$}
 \Text(363,43)[]{$\kbf$}
 \Text(367,68)[]{$\Pbf-\pbf$}
 \Text(315,68)[]{$\Pbf-\pbf+\kbf$}

 \end{picture}
 \caption{The calculation of the e$^+$e$^-\gamma$ wave function.}
\label{photonem}
\end{figure}

Similarly as in Sec. \ref{strucsec} we find that diagram (a) together with a similar diagram in which the photon is emitted by the antifermion
gives the leading contribution in $\alpha$.
Thus at leading order in $\alpha$ the wave function (\ref{photonwf}) is related to $\varphi_\Pbf$ of (\ref{etwfdef}) as shown in Fig. \ref{photonem}.
The analytic expression is in Coulomb gauge \pagebreak
\beqa \label{pwfan}
 \Phi^i_\Pbf(\pbf,\kbf) &=& \frac{i}{E-E_{\pbf-\kbf}-E_{\Pbf-\pbf}-|\kbf|}\frac{1}{2|\kbf|}\left(\delta^{ij}-\frac{k^ik^j}{\kbf^2}\right) \nonumber\\
 &\cdot&\left[\varphi_\Pbf(\pbf)\,ie\gamma^j\Lambda^+(\pbf-\kbf)
 + \Lambda^-(\Pbf-\pbf)\, ie\gamma^j\varphi_\Pbf(\pbf-\kbf) \right] \prm
\eeqa
To find the behavior of $\Phi_\Pbf^i$ for the scales of $\kbf$ and $\qbf$ which are relevant in the bound state, we may use the previous analysis to simplify the expression. Using (\ref{endiff}) and (\ref{crepl}) we have
\beq \label{pwfres}
 \Phi^i_\Pbf(\pbf,\kbf) = \frac{e}{2\kbf^2-2\beta k_\parallel|\kbf|}\left(\frac{P^i}{\mathcal E}- \frac{k^i \beta k_\parallel}{\kbf^2}\right)\left(\varphi_\Pbf(\pbf)-\varphi_\Pbf(\pbf-\kbf)\right)\left[1+\morder{\alpha}\right] \prm
\eeq

To give a probabilistic interpretation for the result (\ref{pwfres}) we need to define the normalization of the wave functions.
We take the normalization of the bound state to be the ``non-relativistic'' one (without a $2\mathcal{E}$ factor) so that
\beq
 \bra{\Pbf' \lambda'}\Pbf \lambda\rangle = (2\pi)^3 \delta^{(3)}(\Pbf'-\Pbf)\delta_{\lambda'\lambda} \prm
\eeq
Then we have
\beq
 1 = \sum_{\lambda'} \int\frac{\ud^3 \Pbf'}{(2\pi)^3}\bra{\Pbf' \lambda'}\Pbf \lambda\rangle = \sum_{\lambda'} \int\frac{\ud^3 \Pbf'}{(2\pi)^3}\bra{\Pbf' \lambda'} \mathbf{1}_{f\bar f} + \mathbf{1}_{f\bar f \gamma} + \cdots \ket{\Pbf \lambda}
\eeq
where the $\mathbf{1}$'s are projection operators on different Fock states\footnote{The contribution from e$^+$e$^-\gamma\rightarrow\gamma$,
which is a higher order correction for a non-relativistic bound state, is not included in (\ref{projres}).
}, \eg,
\beqa \label{projres}
 \mathbf{1}_{f\bar f \gamma} &=& \int\frac{\ud^3 \pbf_1}{(2\pi)^3}\frac{\ud^3 \pbf_2}{(2\pi)^3}\frac{\ud^3 \kbf}{(2\pi)^3}\sum_{s_1,s_2,\lambda}
 c_{s_1}^\dagger(\pbf_1) d_{s_2}^\dagger(\pbf_2)a_\lambda^\dagger(\kbf)\ket{0}\bra{0}c_{s_1}(\pbf_1) d_{s_2}(\pbf_2)a_\lambda(\kbf) \nonumber\\
&=&-\sum_{\alpha,\beta}\int\frac{\ud^3 \pbf_1}{(2\pi)^3}\frac{\ud^3 \pbf_2}{(2\pi)^3}\frac{\ud^3 \kbf}{(2\pi)^3} 2 |\kbf|  \nonumber\\
&\cdot& \widetilde\psi^\dagger_\alpha(\pbf_1) \widetilde\psi_\beta(\pbf_2) \widetilde A^\mu(\kbf)
\ket{0}\bra{0}\widetilde\psi^\dagger_\beta(\pbf_2) \widetilde\psi_\alpha(\pbf_1) \widetilde A_\mu(\kbf) 
\eeqa
in terms of the momentum space field operators defined in (\ref{photonwf}).
Replacing $\bra{0}\rightarrow \bra{\Omega}$ and using translation invariance leads to the correct normalization in the weak coupling limit
\beq \label{normc}
 1 = \int\frac{\ud^3 \pbf}{(2\pi)^3}\trs{\varphi^\dagger_\Pbf(\pbf)\varphi_\Pbf(\pbf)} -
 \int\frac{\ud^3 \pbf}{(2\pi)^3}\frac{\ud^3 \kbf}{(2\pi)^3}\,2|\kbf|\, \trs{\Phi^{\mu\,\dagger}_\Pbf(\pbf,\kbf)\Phi_{\Pbf\,\mu}(\pbf,\kbf)} + \cdots
\eeq
where $\cdots$ stands for the contribution from higher Fock states and the trace is over the Dirac indices. The probability distribution for the Coulomb gauge wave function (\ref{pwfres}) (with $\Phi^0_\Pbf=0$)
is thus
\beqa \label{propd}
 \frac{\ud^6 \mathcal P}{\ud^3 \kbf \ud^3 \qbf}(\qbf,\kbf) &\equiv& \frac{1}{(2 \pi)^6} 2|\kbf| \trs{\Phi^{i\,\dagger}_\Pbf(\Pbf/2+\qbf,\kbf)\Phi_{\Pbf}^i(\Pbf/2+\qbf,\kbf)} \nonumber\\
 &=& \frac{\alpha}{4 \pi^2} \frac{\beta^2\kbf_\perp^2}{|\kbf|^3\left(|\kbf|-\beta k_\parallel\right)^2}\frac{\left|\phi_\Pbf(\qbf)-\phi_\Pbf(\qbf-\kbf)\right|^2}{(2\pi)^3}
\eeqa
expressed in terms of the wave function $\phi_\Pbf$ defined in (\ref{Dirs}), (\ref{hydrwf}).
Inserting the distribution (\ref{propd}) into the normalization condition (\ref{normc}) and using the estimates
(\ref{scales}) we see that the probability for this higher Fock state is \order{\alpha}. 
This is in accord with our above result that $\Delta t_I/\Delta t_F$ is \order{\alpha}: the fraction of time that a 
transverse photon is being exchanged is of \order{\alpha}.
It is straightforward to repeat the analysis for even higher Fock states and check that their probability
is \order{\alpha^n} with $n\ge 2$.
\label{confirm}

\subsection{The photon momentum distribution}

The distribution (\ref{propd}) may be decomposed into three parts
\beqa \label{photonda}
 \frac{\ud^6 \mathcal P}{\ud^3 \kbf \ud^3 \qbf}&=&\frac{\alpha}{4 \pi^2} \frac{\beta^2\kbf_\perp^2}{|\kbf|^3\left(|\kbf|-\beta k_\parallel\right)^2}
 \\&\cdot& \frac{1}{(2\pi)^3} \left[\left|\phi_\Pbf(\qbf)\right|^2 +\left|\phi_\Pbf(\qbf-\kbf)\right|^2 -2\, \mathrm{Re}\left\{\phi_\Pbf(\qbf)\phi^*_\Pbf(\qbf-\kbf)\right\}\right] \prm \nonumber
\eeqa
In the square of the wavefunction (\ref{pwfres}) the first two terms correspond to graphs where the photon is emitted and absorbed by the
same particle (see Fig. \ref{photonem}), and the interference comes from photon exchange. The factor which multiplies
the two-particle wave functions in (\ref{photonda}) depends only on $\kbf$ and
is basically the distribution of photons emitted by a single fermion. As one might expect, each of the three terms is separately
infra-red divergent as $|\kbf|\rightarrow 0$, but the sum is infra-red safe as a consequence of the charge neutrality of the system.
However, there is an ultra-violet divergence as $|\kbf|\rightarrow \infty$ which stems from the first two terms. For high $\kbf$ the 
photon wavelength is short, and its emission is incoherent. The divergence
reflects the distribution of photons emitted from a free electron.

Above we have used an asymmetric coordinate convention because of notational simplicity. To better understand the distribution, we use
in this section a more natural, symmetric convention, where the momenta of the fermion, the antifermion and the photon are
$\Pbf/2+\qbf-\kbf/2$, $\Pbf/2-\qbf-\kbf/2$ and $\kbf$, respectively. That is, we shift
$\qbf \rightarrow \qbf +\kbf/2$ and thus redefine
\beq
 \qbf \equiv \pbf - \Pbf/2 - \kbf/2
\eeq
in the presence of a photon.
We furthermore define the Lorentz contracted momentum variables
\beqa \label{lcvar}
 \hat \kbf &=& (\hat \kbf_\perp,\hat k_\parallel) \equiv (\kbf_\perp,k_\parallel/\gamma) \nonumber\\
 \hat \qbf &=& (\hat \qbf_\perp,\hat q_\parallel) \equiv (\qbf_\perp,q_\parallel/\gamma) \prm
\eeqa
In terms of these variables the distribution (\ref{propd}) becomes [remembering (\ref{hydrwf})]
\beqa
 \frac{\ud^6 \mathcal P}{\ud^3 \hat \kbf \ud^3 \hat \qbf}  &=& \frac{\alpha}{4 \pi^2} \frac{\gamma\beta^2\hat \kbf_\perp^2}{(\hat \kbf_\perp^2+\gamma^2 \hat k_\parallel^2)^{3/2}\left(\sqrt{\hat \kbf_\perp^2 + \gamma^2 \hat k_\parallel^2}-\beta \gamma \hat k_\parallel\right)^2} \nonumber\\
&\cdot& \frac{\left|\phi_{CM}(\hat \qbf+\hat\kbf/2)-\phi_{CM}(\hat \qbf- \hat \kbf/2)\right|^2}{(2\pi)^3} \prm
\eeqa
The azimuthally averaged photon distribution, integrated also over the relative momentum $\hat \qbf$ of the fermions, reads
\beqa \label{intd}
 \frac{\ud^2 \mathcal P}{\ud \hat k\, \ud\! \cos \theta} &\equiv&  \hat k^2 \int_0^{2 \pi} \ud \varphi \int \ud^3 \hat \qbf \frac{\ud^6 \mathcal P}{\ud \hat \kbf^3 \ud \hat \qbf^3} \nonumber\\
 &=& \frac{\alpha}{4 \pi^2} \frac{\gamma\beta^2 (1-\cos^2 \theta)}{(1+\beta^2\gamma^2\cos^2\theta)^{3/2}\left(\sqrt{1+\beta^2\gamma^2\cos^2\theta}-\beta \gamma \cos \theta \right)^2} \nonumber\\
&\cdot&\frac{1}{\hat k}\int_0^{2\pi}\ud \varphi \int \frac{\ud^3\hat \qbf}{(2\pi)^3} \left|\phi_{CM}(\hat \qbf+\hat\kbf/2)-\phi_{CM}(\hat \qbf- \hat \kbf/2)\right|^2
\eeqa
where $\theta$ is the angle between $\Pbf$ and $\hat \kbf$, $\varphi$ is the remaining azimuthal angle and $\hat k=|\hat \kbf|$. Note that the frame dependence
only appears in the angular distribution multiplying the integrals. An exactly Lorentz contracting distribution would be completely frame independent
when expressed in terms of the variables (\ref{lcvar}). The angular distribution is the same as that of a free particle, $\textrm{e}\rightarrow \textrm{e}+\gamma$.

\subsection{The photon distribution for the ground state}

The photon distribution (\ref{intd}) is valid for any e$^+$e$^-$ bound state wave function $\phi_{CM}$. Let us study the ground state
\beq \label{gswf}
 \phi_{CM}^{(0)}(\hat \qbf) = \sqrt{\frac{512 \pi}{\alpha^3 m^3}}\frac{1}{\left[1+\frac{\hat \qbf^2}{\left(\alpha m / 2\right)^2}\right]^2} \prm
\eeq
For the wave function (\ref{gswf})  the integral over $\hat \qbf$ in (\ref{intd}) is independent of the angles $\theta$ and $\varphi$. Thus the $\theta$ and $\hat k$ dependence
of the distribution completely factorizes, \ie,
\beq
 \frac{\ud^2 \mathcal P}{\ud \hat k\, \ud\! \cos \theta} = \frac{\alpha}{2 \pi} f(\cos\theta) g(\hat k)
\eeq
where the angular dependence is
\beq \label{fdef}
 f(\cos\theta) \equiv \frac{\gamma\beta^2 (1-\cos^2 \theta)}{(1+\beta^2\gamma^2\cos^2\theta)^{3/2}\left(\sqrt{1+\beta^2\gamma^2\cos^2\theta}-\beta \gamma \cos \theta \right)^2} 
\eeq
and
\beq \label{gdef}
 g(\hat k) \equiv \frac{1}{\hat k}\int \frac{\ud^3 \qbf}{(2\pi)^3} \left|\phi_{CM}^{(0)}(\hat \qbf+\hat\kbf/2)-\phi_{CM}^{(0)}(\hat \qbf- \hat \kbf/2)\right|^2 \prm
\eeq

\begin{figure}
 \centering
 \includegraphics[width=\textwidth]{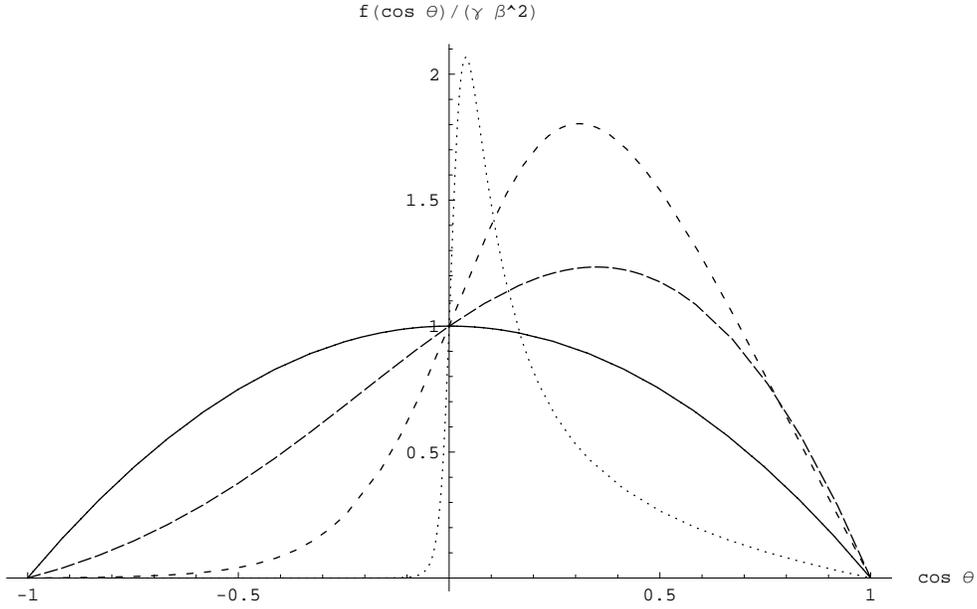}
 \caption{The angular dependence the contracted and integrated photon distribution (\ref{intd}) in the positronium ground state. The lines show the  angular distribution $f(\cos \theta)/(\gamma\beta^2)$ [defined in (\ref{fdef})] for $\beta=0.001$, $0.5$, $0.9$ and $0.999$. For $\beta=0.001$ (solid line) the distribution is close to the symmetric limit (\ref{smallb}). For $\beta=0.999$ (dotted line) the distribution approaches the limit (\ref{largeb}).}
\label{angplot}
\end{figure}

The dependence of the function $f$ on $\gamma$ in (\ref{fdef}) gives the deviation from exact Lorentz contraction. The angular dependence is plotted for various values of $\gamma$ in Fig. \ref{angplot}.
In particular, we have
\beqa \label{smallb}
 \beta^{-2} f(\cos\theta) &\longrightarrow& (1-\cos^2\theta) \quad\textrm{as}\ \beta\rightarrow 0 \\
 f(\cos \theta) &\longrightarrow& 4\,\Theta(\cos \theta)\,\frac{1-\cos^2\theta}{\cos\theta}  \quad\textrm{as}\ \beta\rightarrow 1 \label{largeb}
\eeqa 
where $\Theta$ is the step function. Thus for large boosts $\gamma \gg 1$ almost all of the photons have $k_\parallel > 0$.

Let us study the (frame independent) radial distribution $g(\hat k)$ of (\ref{gdef}). For $\hat k\gg \alpha m$, the interference term is negligible and we have
\beq 
  g(\hat k) \simeq \frac{2}{\hat k} 
\eeq
so that the distribution falls as $1/\hat k$ for large $\hat k$.
For $\hat k \ll \alpha m$ we have
\beq
 g(\hat k) \simeq \frac{1}{\hat k}\int \frac{\ud^3 \hat \qbf}{(2\pi)^3} \left|\hat \kbf \cdot \nabla \phi_{CM}^{(0)}(\hat \qbf)\right|^2
 = \frac{4 \hat k}{(\alpha m)^2}
\eeq
which reflects the decoupling of long wavelength photons from neutral positronium.
The limit of small and large $\hat k$ behaviors
found here is similar for excited states of positronium.

\begin{figure}
 \includegraphics[width=\textwidth]{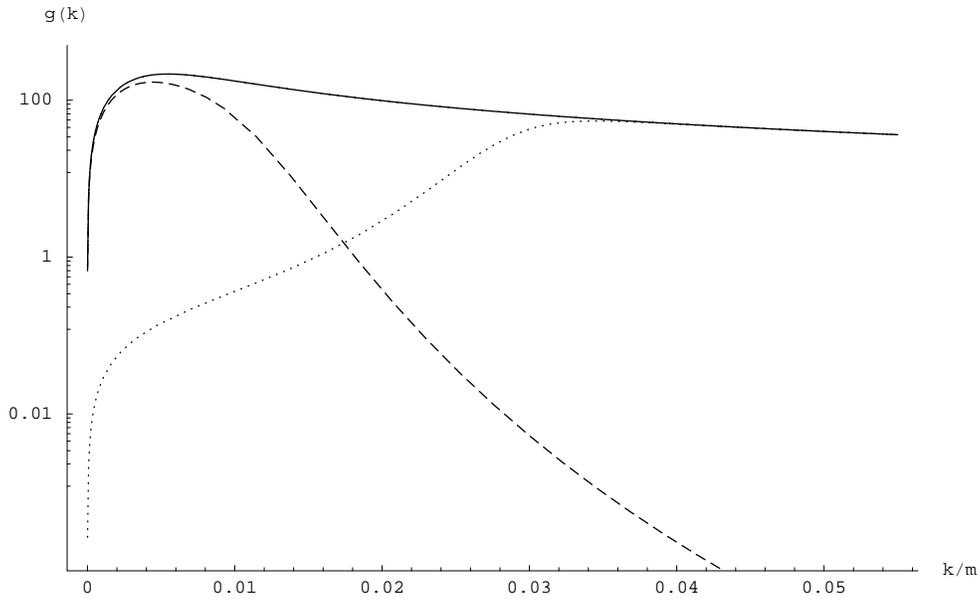}
 \caption{The radial behavior for the integrated photon distribution (\ref{intd}). Solid line is the function $g(\hat k)$ defined in (\ref{gdef}). Dashed and dotted lines
 are the contributions to $g(\hat k)$ from the regions $\hat q < 0.7 \alpha m$ and $\hat q > 2 \alpha m$, respectively.}
\label{radplot}
\end{figure}

For general values of $\hat k$ the distribution $g(\hat k)$ may be found numerically. It is also possible to study the correlation between
the magnitudes of $\hat q$ and $\hat k$ by including a $\hat q$ dependent cut in the integration of (\ref{gdef}). This is illustrated 
in Fig. \ref{radplot}. Almost all of the long wavelength photons $\hat k\ll\alpha m$ are present when the size of the fermion system (given by $1/\hat q$)
is also large. Conversely, soft photon emission by compact fermion systems is suppressed.

\begin{figure}
 \includegraphics[width=\textwidth]{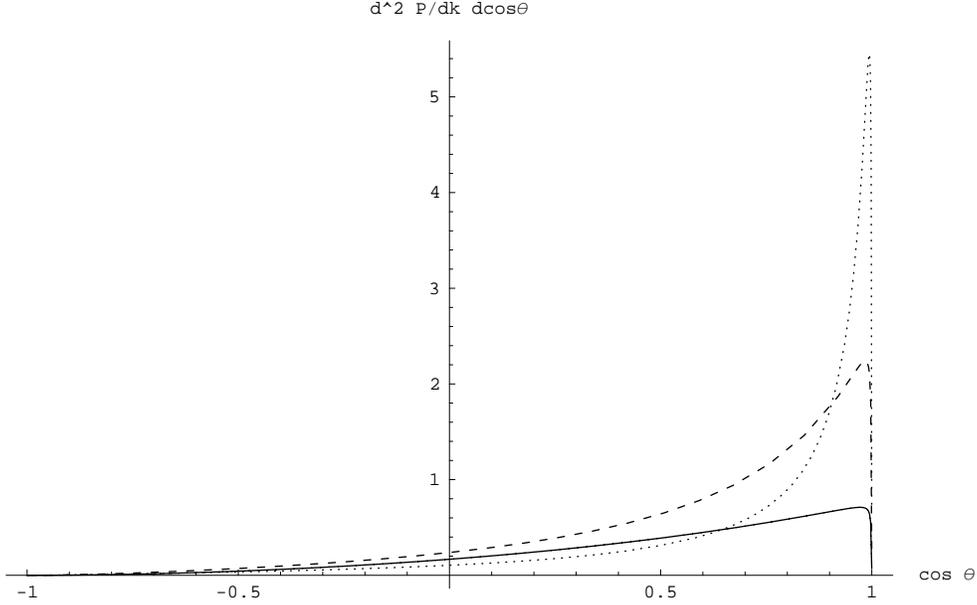}
 \caption{The angular dependence of the integrated photon distribution (\ref{intdnc}) for a system with velocity $\beta=0.999$. Solid, dashed and dotted lines show the angular distribution for $k=0.3 \alpha m$, $\alpha m$ and $3 \alpha m$, respectively. }
\label{angplot2}
\end{figure}

It is also instructive to plot the photon distribution in terms of the usual (uncontracted) momentum variable $\kbf$. In terms of $\kbf$ the integrated photon distribution (\ref{intd}) for the ground state becomes
\beqa \label{intdnc}
 \frac{\ud^2 \mathcal P}{\ud k\, \ud\! \cos \theta} 
 &=& \frac{\alpha}{2 \pi} \frac{\beta^2 (1-\cos^2 \theta)}{\left(1-\beta \cos \theta \right)^2} \nonumber\\
&\cdot&\frac{1}{k} \int \frac{\ud^3\hat \qbf}{(2\pi)^3} \left|\phi_{CM}^{(0)}(\hat \qbf+\hat\kbf/2)-\phi_{CM}^{(0)}(\hat \qbf- \hat \kbf/2)\right|^2 
\eeqa
where the integral only depends on $\hat k = k \sqrt{1 - \beta^2 \cos^2\theta}$. For $\beta \rightarrow 0$ we have $\kbf=\hat \kbf$ and the distribution (\ref{intdnc})
coincides with the $\hat\kbf$ distribution described above. However, for large boosts there is a big difference between the distributions. We plot the angular distribution
for different values of $k$ in Fig. \ref{angplot2} for a system with $\beta=0.999$. There is a drastic difference when compared to the contracted distribution of Fig. \ref{angplot}. Due to the Lorentz contraction effect, the distribution is peaked at forward angles, in particular for large $k$.

In our calculation (\ref{pwfres}) we assumed that
$\hat k \sim \alpha m$. Hence the results do not necessarily hold for arbitrarily large values of $\hat k$. However, it is straightforward to check
that the $1/k$ behavior is consistent with the small $z$ behavior $\sim 1/z$ of the Weizs\"acker-Williams photon distribution
function
\beq
 f_\gamma(z) = \frac{\alpha}{2\pi}\left[\frac{1+(1-z)^2}{z}\right]\log\frac{s}{m_e^2}
\eeq
which gives the probability of finding a physical, transverse photon with a longitudinal momentum fraction $z$ in a high energy electron (see, \eg, \cite{Peskin}).

\subsection{High $|\Pbf|$ limit}

Finally we check that our result for the photon distribution (\ref{propd}) is compatible with the light-front results.
It is generally accepted that the usual equal-time picture of quantum field theory in the infinite momentum frame
coincides with the one calculated at equal light-front time $x^+=t+z=0$. Thus to compare with the light-front results, we need to take the limit $|\Pbf|\rightarrow \infty$.
It is easy to check that in this limit our solutions for the e$^+$e$^-$ wave function (\ref{hydrwf}) reduce to the light-front wave functions
at leading order in $\alpha$ (see \cite{Burkardt,Lepage}),
\beq \label{wflim}
 \phi_\Pbf(\qbf) = \sqrt{\frac{(2\pi)^3}{|\Pbf|}}\phi_{LF}(y,\qbf_\perp)\left[1 + \morder{\frac{m}{|\Pbf|}}\right]\crm
\eeq
where $y$ is the momentum fraction $y\equiv p_\parallel/|\Pbf| \simeq 1/2 + q_\parallel/2\gamma m$. 
As the two-particle Fock state is leading in $\alpha$ also on the light front the normalization can be taken to be
$1 = \int \ud \pbf_\perp \int_0^1 \ud y |\phi_{LF}(y,\pbf_\perp)|^2$.

The light-front photon distribution is calculated in \cite{Burkardt}. 
As in our calculation the probability of the Fock state with one additional photon is \order{\alpha}.
To compare with our result we need the relations
\beqa \label{klim}
 |\kbf| &=& |\Pbf||x|\left[1 + \morder{\frac{m}{|\Pbf|}}\right] \nonumber\\
 \frac{1}{|\kbf|-\beta k_\parallel} &=& \Theta(x)\frac{2|\Pbf|x}{(2mx)^2+\kbf_\perp^2}\left[1 + \morder{\frac{m}{|\Pbf|}}\right] \prm
\eeqa
where $x$ is the momentum fraction $x\equiv k_\parallel/|\Pbf| \simeq k_\parallel/2\gamma m$. Note that (\ref{scales}) gives
\beq
 x \sim \alpha \sim \left(y-\frac 12\right) \prm
\eeq
Inserting (\ref{wflim}) and (\ref{klim}) into (\ref{propd}) we find
\beqa \label{lflimit}
 \Pbf^2\frac{\ud^6 \mathcal P}{\ud^3 \kbf \ud^3 \qbf}(\qbf,\kbf)\! &\longrightarrow&\!
 \frac{\ud^6 \mathcal P}{\ud y \ud^2 \kbf_\perp \ud x\ud^2 \qbf_\perp}(x,\kbf_\perp,y,\qbf_\perp)\\ \nonumber
 &=&\!\! \frac{\alpha}{\pi^2}\Theta(x)\frac{\kbf_\perp^2/x}{\left[(2mx)^2+\kbf_\perp^2\right]^2}\left|\phi_{LF}(y,\qbf_\perp)-\phi_{LF}(y-x,\qbf_\perp-\kbf_\perp)\right|^2
\eeqa
which agrees with the light-front result \cite{Burkardt}\footnote{The result (\ref{lflimit}) is missing the constraints $x<y<1$ which appear at the light front.
They are unimportant in the weak coupling limit as $x\sim\alpha$ and the light-front wave functions are peaked at $y \simeq 1/2$.}.

\section{Summary and discussion}

We studied the frame dependence of non-relativistic QED bound states such as positronium or the hydrogen atom. 
Starting from the exact field theoretical Bethe-Salpeter equation we evaluated the wave function to leading order in $\alpha$ in all Lorentz frames. 
Using a time-ordered formalism we confirmed
the expected result: the e$^+$e$^-$ equal-time wave functions exactly Lorentz contract in boosts while the mass spectrum is invariant.

We also solved the leading component of the wave function of the e$^+$e$^-\gamma$ Fock state. We saw that this Fock state was a next-to-leading correction
with a probability \order{\alpha}. The resulting photon distribution did not contract exactly in boosts, similarly to the radiation from a single electron,
$\textrm{e}\rightarrow \textrm{e}+\gamma$. The infinitely boosted limit of the distribution was seen to coincide with light-front results.

The ``old-fashioned'' time-ordered approach is natural when considering wave functions evaluated at equal time. The advantages of the time-ordered formalism for bound states are well known from studies of positronium in the center-of-mass frame -- see, \eg, \cite{Labelle} and references therein.
In our case the time-ordering was  helpful in the analysis of the Fock state structure of the bound state and in determining the correct order of $\alpha$ for various interaction kernels. 

The covariant (four-dimensional) form of the Bethe-Salpeter equation  
determines also the dependence on the relative time $t$ of the bound  
state constituents. The time dependent wave function $\Psi(t,\xbf)$ may  
be derived at weak coupling as in the 1+1 dimensional case studied in  
section II.C of \cite{Jarv}. The result is\footnote{The time dependence  
in the case of non-relativistic center-of-mass motion was given in  
\cite{Bethe,Hayashi}.}
\beq \label{reltdep}
\Psi_\Pbf(t,\xbf) = \exp\left(-\frac{i\alpha  
|t|}{2\gamma\sqrt{\xbf_\perp^2 + \gamma^2 \tilde x_\parallel^2}}\right)  
\tilde \varphi_\Pbf(\tilde x_\parallel,\xbf_\perp)
\eeq
where $\tilde x_\parallel \equiv x_\parallel-\beta t$ is the  
longitudinal distance between the constituents adjusted for the  
displacement $\beta t = |\Pbf|t/E$ of the center of mass, and $\tilde  
\varphi_\Pbf$ is the equal-time wave function $\varphi_\Pbf$ of  
(\ref{Dirs})  Fourier transformed to coordinate space.
Combining (\ref{reltdep}) with our result (\ref{Dirs}) one may check  
that the covariant Lorentz transformation formula \cite{Brodsky} of the  
e$^+$e$^-$ wave function is satisfied,
\beq
\Psi_{\Pbf'}(t',\xbf') = S(\Lambda) \Psi_{\Pbf=0}(t,\xbf) S^{-1}(\Lambda)
\eeq
where $S(\Lambda)$ is the standard spin transformation matrix of the Dirac equation. 

It would be interesting to extend the  
analysis presented here to the next-to-leading order in $\alpha$ and check if the correction
to the wave function contracts classically.
Likewise one could consider the frame dependence of non-relativistic  
bound states in other theories, \eg, for states bound by scalar  
exchange.

In previous work scant attention has been paid to the description of moving bound states in field theory, 
even though much is known about the center-of-mass frame solutions, in particular for non-relativistic systems. 
However, the problem is non-trivial and certainly worth studying.
The usual center-of-mass equal-time wave functions are related by an infinite boost to the light-front wave functions which appear in the parton model.
Hence a better understanding of the behavior of (relativistic) bound states under boosts could help to relate the non-relativistic quark model and the
parton model of hadrons. Evaluating the boosted wave functions of the simpler, non-relativistic systems which we studied here is a first step towards understanding the boost properties of relativistic systems such as hadrons.

\subsection*{Acknowledgments}

I would like to thank Paul Hoyer for suggesting this topic and for several useful discussions.
I am also grateful for Paul Hoyer and Stanley Brodsky for comments on the paper.

\end{document}